\documentstyle[12pt,epsfig]{article}
\author{Juli\'an Candia$^{a}$ and Ezequiel V. Albano$^{b}$\\{}\\
$^a${\small\it Departamento de F\'{\i}sica, UNLP, 
CC67, 1900 La Plata, Argentina}\\
$^b${\small\it Instituto de Investigaciones Fisicoqu\'{\i}micas
Te\'{o}ricas y Aplicadas}\\{\small\it (INIFTA), UNLP, CONICET, 
Suc.4, CC16, 1900 La Plata, Argentina}}

\title{Comparative study of an Eden model 
for the irreversible growth of spins and the equilibrium 
Ising model}
\begin{document}
\maketitle

\begin{abstract}
The Magnetic Eden Model (MEM) [N. Vandewalle et al., 
Phys. Rev. E. {\bf 50}, R635 (1994)]
with ferromagnetic interactions between
nearest-neighbor spins is studied in $(d+1)-$dimensional rectangular
geometries for $d = 1,2$. In the MEM, magnetic clusters are grown
by adding spins at the boundaries of the clusters. 
The orientation of the added spins
depends on both the energetic interaction with already deposited spins and the 
temperature, through a Boltzmann factor. 
A numerical Monte Carlo investigation of the MEM
has been performed and the results of the simulations have been analyzed 
using finite-size scaling arguments. 
As in the case of the Ising model, the MEM in $d = 1 $ is non-critical 
(only exhibits an ordered phase at $T= 0$). 
In $d = 2$ the MEM exhibits an order-disorder transition of second-order
at a finite temperature. Such transition has been characterized
in detail and the 
relevant critical exponents have been determined. These exponents are
in agreement (within error bars) with those of the Ising model in 
2 dimensions. Further similarities between
both models have been found by evaluating the probability distribution 
of the order parameter, the magnetization and the susceptibility. 
Results obtained by means of extensive computer simulations allow us 
to put forward a conjecture which  establishes
a nontrivial correspondence between the MEM
for the irreversible
growth of spins and the equilibrium Ising model. This conjecture
is certainly a theoretical challenge and its
confirmation will contribute to the development
of a framework for the study of irreversible
growth processes.
\end{abstract}

\section{Introduction}

The study of kinetic growth models such as
directed percolation, Eden growth, ballistic
deposition, diffusion limited aggregation, random
deposition with and without relaxation, cluster-cluster
aggregation, etc., is motivated by their interest in
many areas of scientific research and technology such
as polymer science, crystal and polycrystalline growth,
gelation, fracture propagation, epidemic spreading,
bacterial and fungi growth colonies, 
colloids, etc. \cite{fam,bar,shl1,shl2,mar}.
Within this context the Eden model \cite{ede} has become an
archetype growth model. Eden clusters are compact but 
the self-affinity that characterizes the behavior
of the growing interface is of much interest 
(see e.g. \cite{ed1,ed2,ed3,ed4,ed5,ed6}). 
Few years ago  Ausloos et al. \cite{mem1} 
have introduced an additional degree
of freedom to the Eden model, namely the spin of the added particles.
More recently, the Eden growth of clusters of charged 
particles has also been studied \cite{qe}. 
 
In the Magnetic Eden Model (MEM) \cite{mem1} 
with spins having two 
orientations (up and down) the growth of the cluster starts
from a single seed, e.g.
a spin up seed, placed at the center of the two-dimensional square lattice,
whose sites are labelled by their rectangular coordinates $(i,j)$.
Then, the growth process of the resulting magnetic cluster  
consists in adding further spins to the growing cluster
taking into account the corresponding interaction energies. By analogy
to the Ising model \cite{ising} one takes $J$ 
as the coupling constant between
nearest-neighbor (NN) spins $S_{ij}$ and the energy $E$ is then given by
\begin{equation}
E = - \frac{J}{2} \sum_
{ \langle ij,i^{'}j^{'}  \rangle } S_{ij}S_{i^{'}j^{'}} ,   
\end{equation}
where $\langle ij,i^{'}j^{'} \rangle$ means 
that the summation is 
taken over occupied NN sites.
The spins can assume two values, namely $S_{ij}= \pm 1$.
Throughout this work we set the Boltzmann constant equal
to unity ($k_{B} \equiv 1$), we
consider $J > 0$
(i.e., the ferromagnetic case) and we take the
absolute temperature $T$ measured in units of $J$.
In the MEM a spin is added to the cluster with a probability
proportional to the Boltzmann factor exp$(-\Delta E /T)$, 
where $\Delta E$ is the total energy change involved. 
It should be noted that at each step all sites of
perimeter are considered and the probabilities of adding up
and down spins have to be evaluated. After proper normalization
of the probabilities the growing site and the orientation of the
spin are determined through a pseudo-random number generator.
 
It is worth mentioning that the MEM has originally been motivated 
by the study of the structural properties of magnetically 
textured materials \cite{mem1}. While  these previous studies of the 
MEM were mainly devoted to determine the
lacunarity exponent and the fractal dimension of the set of parallel
oriented spins \cite{mem1}, the aim of the 
present work is to complement these
previous investigations by studying the critical behavior of the
MEM using extensive Monte Carlo simulations and applying a
finite-size scaling theory. Also, our study is performed
in confined (stripped) geometries which resemble
recent experiments where the growth of 
quasi-one-dimensional strips of Fe on a Cu(111) vicinal surface
\cite{iron} and Fe on a W(110) stepped substratum \cite{w110}  
have been performed. In fact, the preparation and
characterization of magnetized nanowires is of great interest
for the development of advanced 
microelectronic devices \cite{iron,w110,nw1,nw2}.  
Furthermore, the growth of metallic multilayers  
of Ni and Co separated by a Cu spacer
layer has recently been also studied \cite{cobre}.

Another goal of the present work is to compare the 
results obtained for the MEM with the 
well known behavior of the classical Ising model \cite{ising,Wu}, 
an archetypical  
model in the study of thermally driven (reversible)
phase transitions in equilibrium systems. The Ising 
Hamiltonian $(\bf{H})$ is given by
\begin{equation}
{\bf H} = - \frac{J}{2} \sum_
{ \langle ij,i^{'}j^{'} \rangle} S_{ij}S_{i^{'}j^{'}} \ \ ,   
\end{equation}
where  $\langle ij,i^{'}j^{'} \rangle$ means 
that the summation runs over all NN sites,
$S_{ij} = \pm 1$ is the state of the spin at the site
of coordinates $(i,j)$ and $J$ is the 
coupling constant ( $J > 0$).

The MEM is also similar to a family of models for the stochastic growth
of crystals generically known as crystal growth models (CGM)
\cite{x,y,m,d,w}, for a review see e.g. \cite{q}.
As in the MEM, in the case of CGM each atom is adsorbed with 
a given probability conditional to the actual configuration of neighboring
atoms on the previous layer(s). However, in contrast to the MEM, the 
crystal is supposed to grow layer after layer. 
It should also be noticed that 
relationships established between CGM and a special class 
of Ising models \cite{m,w,z} have allowed to derive exact results.
Therefore, useful comparisons with the MEM will be also discussed
in the presentation of our results.

This paper is organized as follows: in Section 2
we give details on the simulation method, Section 3 is devoted
to the presentation and discussion of the results obtained for the
MEM in $(1 + 1)$-dimensions, while Section 4 refers to results 
corresponding to $(2 + 1)$-dimensions. In Sections 3 and 4,
detailed discussions comparing our results with the behavior of the 
Ising magnet are outlined.
Finally our conclusions are stated in Section 5.

\section{Description of the simulation method}

The MEM in $(1 + 1)-$dimensions is studied in the square lattice using a
rectangular geometry $L \times M$ with $M \gg L$ 
and imposing periodic boundary
conditions along the $L-$direction. The location of each site on the
lattice is specified through its rectangular coordinates $(i,j)$,
($1 \leq i \leq M$, $1 \leq j \leq L)$.
The starting seed for the growing cluster is a column
of parallel oriented spins placed at $i=1$. 
It should be noticed that previous simulations
of the MEM were restricted to rather modest cluster sizes, i.e.
containing up to 8000 spins \cite{mem1}, while in the present
work clusters having up to $10^{9}$ spins have been typically grown.
We have also studied the MEM in $(2+1)-$dimensions employing
a $L \times L \times M$ geometry ($M \gg L$) with periodic 
boundary conditions along both $L-$directions.

\section{Study of the MEM in $(1+1)-$dimensions: results and discussion}

Magnetic Eden clusters grown on a stripped geometry of
finite linear dimension $L$ at low temperatures
show an interesting behavior that we call magnetization reversal.
In fact, we have observed that long clusters are constituted
by a sequence of well ordered magnetic domains. Spins belonging
to each domain, of average length $l_{D} \gg L$, have mostly the same
orientation and consecutive domains have opposite orientation.
Let $l_{R}$ be the characteristic length for the occurrence
of the magnetization reversal. Since $l_{R}\sim  L$, 
we then conclude that the problem has two characteristic 
length scales, namely $l_{D}$ and $l_{R}$,
such that $l_{D} \gg l_{R}$. 

In ordinary thermally driven phase transitions, the system
changes from a disordered state at high temperatures to a
spontaneously ordered state at temperatures below some critical
value $T_{c}$ where a second-order phase transition takes place.
Regarding the Ising model,
one has that, in the absence of an external magnetic
field ($H=0$), the low temperature ordered phase is a state with
non-vanishing spontaneous magnetization  ($ \pm M_{sp}$).
This spontaneous symmetry breaking is possible
in the thermodynamic limit only. In fact,
it is found that the magnetization $M$ of a finite sample 
can pass with a finite probability from a value near
$+M_{sp}$ to another
near $-M_{sp}$, as well as in the opposite direction. Consequently,
the magnetization of a finite system, averaged over a
sufficiently large observation time, 
vanishes at every positive temperature.
The equation $M(T,H=0) \approx 0$ holds if the observation time ($t_{obs}$)
becomes larger than the ergodic time ($t_{erg}$), which is defined as
the time needed to observe the system passing from $\pm M_{sp}$ 
to $\mp M_{sp}$. 
Increasing the size of the sample the ergodic time increases too,
such that in the thermodynamic limit ergodicity is
broken due to the divergence of the ergodic time, yielding
broken symmetry.
Since Monte Carlo simulations are restricted
to finite samples, the standard procedure to avoid the problems
treated in the foregoing discussion is to consider the
absolute magnetization as an order parameter \cite{kuku}.
Turning back to the MEM, we find  that the phenomenon of magnetization
reversal also causes the
magnetization of the whole cluster to vanish at every non-zero
temperature, provided that the length of the cluster $l_{C}$ (which plays the 
role of  $t_{obs}$) is much larger than $l_{D}$ (which plays the role
of $t_{erg}$ ).
Therefore, we have measured the mean absolute column magnetization, given by
\begin{equation}
|m(i,L,T)| = \frac{1}{L} |\sum_{j = 1}^{L} S_{ij}| \ \ .                       
\end{equation}

In the stripped geometry used in this work the bias introduced
by the lineal seed (a starting column made up entirely of up spins)
can be avoided by calculating relevant properties after 
disregarding spins within a distance approximately equal 
to few times $L$ from the seed.
The procedure of column averaging out from the
transient region represents a significant advantage
of the stripped geometry used for the
simulation of the MEM. In fact, when a single seed 
at the center of the sample is used, the definition
of the average magnetization of the whole cluster 
is strongly biased by the cluster's
kernel orientation at the early stages of the growing process.
In addition, using several randomly generated seeds
we could also establish that the system evolves into
a given stationary state independently of the seed employed.

\begin{figure}
\centerline{{\epsfxsize=4.0in\epsfysize=3.2in \epsffile{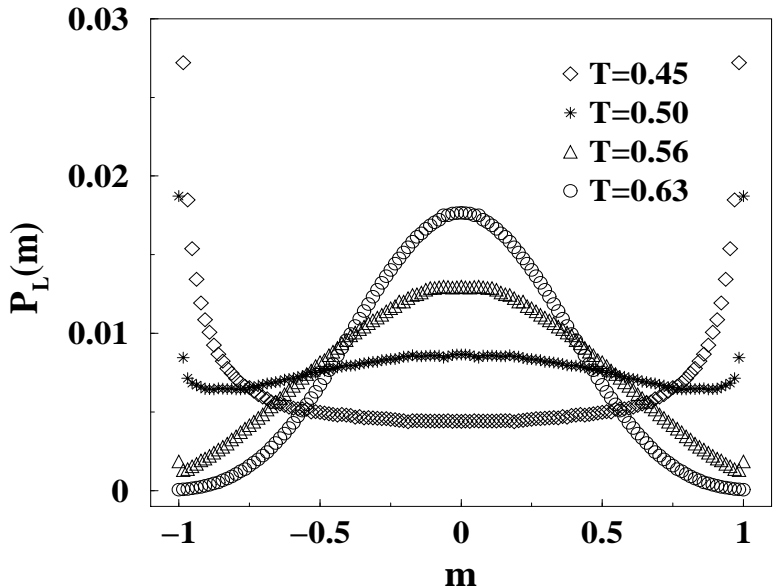}}}
\caption{Plots of the probability distribution of the mean
column magnetization $P_{L}(m)$ versus $m$ for the fixed lattice
width $L=128$ and different temperatures, as indicated in
the figure. The sharp peaks at $m = \pm 1$ for $T=0.45$ have been
truncated, in order to allow a detailed observation of the plots
corresponding to higher temperatures.
This behavior resembles that of the one-dimensional
Ising model. More details in the text.}
\label{fig1}
\end{figure}         

The mean column magnetization
is a fluctuating quantity that can assume $L+1$ values. 
Then, for given values of both $L$ and $T$,
the probability distribution of the mean column
magnetization $(P_{L}(m))$ 
can be evaluated, since it represents the
normalized histogram of $m$ taken over a sufficiently large number
of columns in the stationary region \cite{alfa,beta,gamma}.
In the thermodynamic limit
the probability distribution $(P_{\infty}(m))$
of the order parameter of an equilibrium system
at criticality is universal (up to rescaling of the 
order parameter) and thus it contains
very useful and interesting information
on the universality class of the system \cite{pd1,pd2,pd3}. 
Figure 1 shows the thermal dependence of $P_{L}(m)$
for a fixed lattice size ($L=128$)
as obtained for the MEM.
At high temperatures $P_{L}(m)$ is a Gaussian
centered at $m=0$ but when the temperature gets lowered, the distribution
broadens and develops two peaks at $m=1$ and $m=-1$.
Further decreasing the temperature
causes these peaks to become dominant
while the distribution turns distinctly non-Gaussian, exhibiting a
minimum just at $m=0$. The emergence of the maxima
at $m= \pm 1$ is quite abrupt.
This behavior reminds us the order parameter probability
distribution characteristic of the one dimensional Ising model.
In fact, for the well studied $d-$dimensional Ising model \cite{gamma,lali}, 
we know that
for $T > T_{c}$, $P_{L}(M)$ is a Gaussian centered at $M=0$, given by
\begin{equation}
P_{L}(M) \propto \exp \left({-M^{2}L^{d}} \over {2T \chi} \right) \ \    ,    
\end{equation}
where the susceptibility $\chi$ is related to
order parameter fluctuations by
\begin{equation}
\chi = \frac {L^{d}}{T} \left(\langle M^{2} \rangle - \langle M \rangle^{2} \right)  .                 
\end{equation} 

\begin{figure}
\centerline{{\epsfxsize=4.0in\epsfysize=3.0in \epsffile{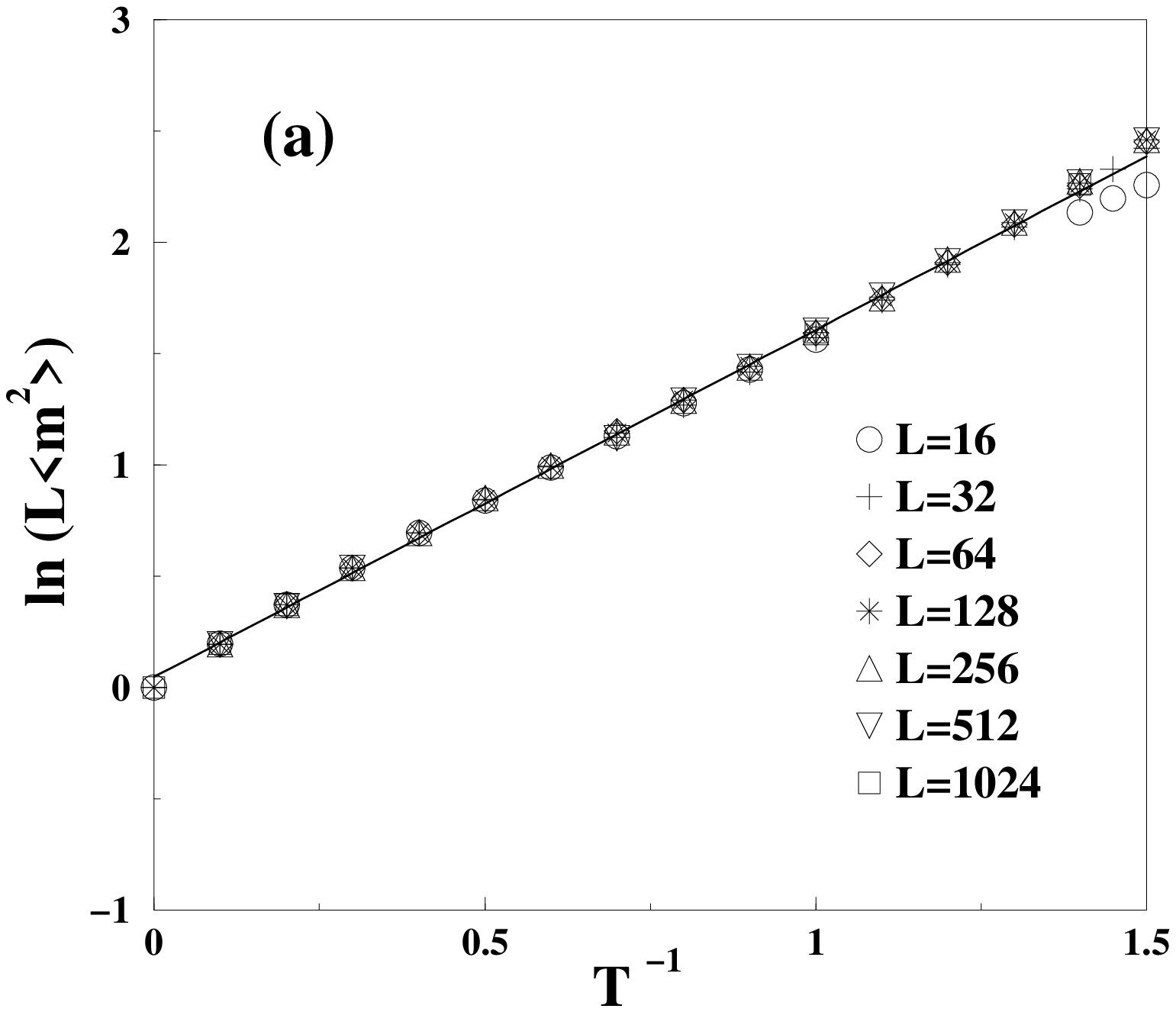}}}
\centerline{{\epsfxsize=4.0in\epsfysize=3.0in \epsffile{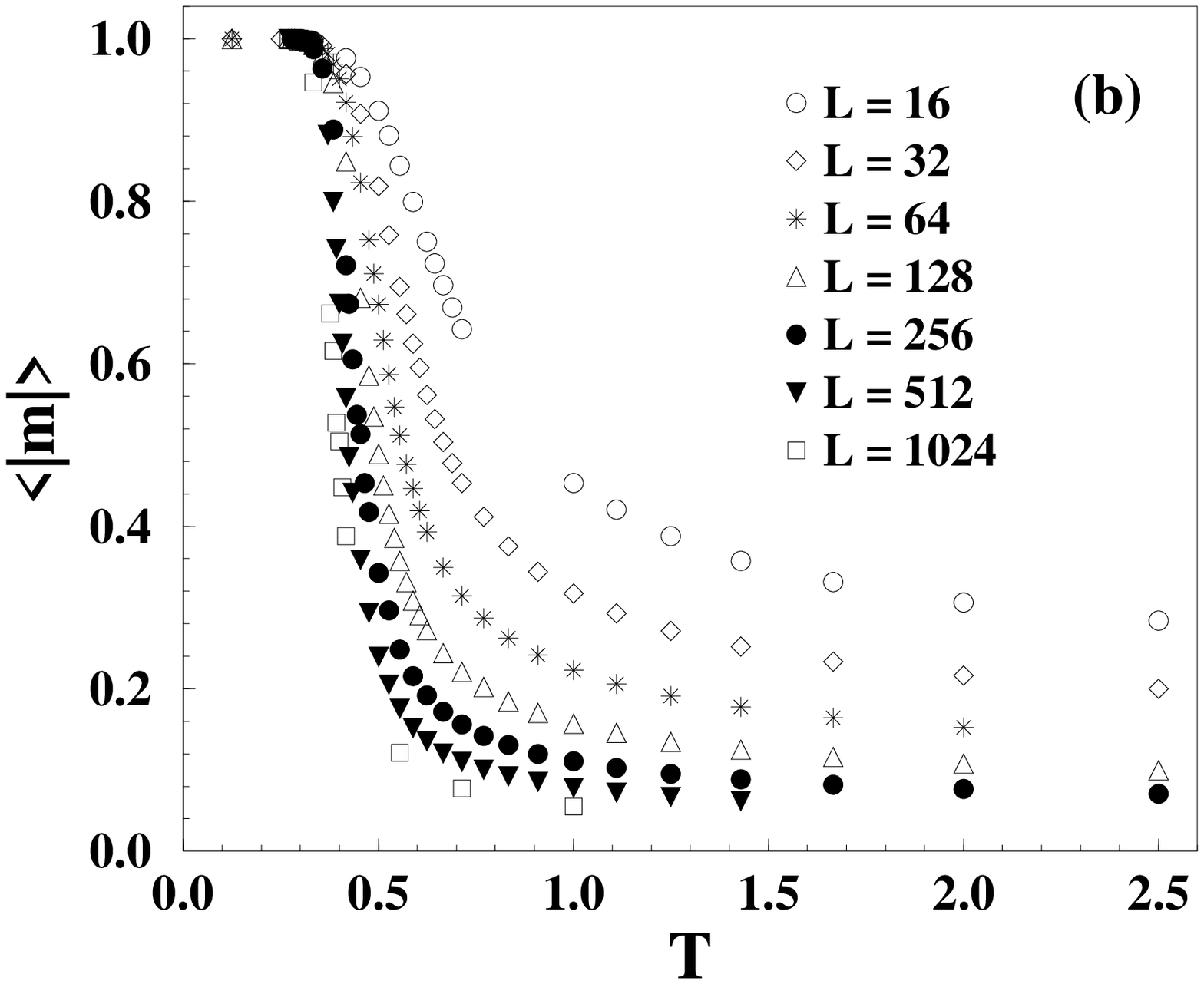}}}
\caption{Data for strip widths in the range $16 \leq L \leq 1024$
as indicated in the figures.
(a) ln-linear plots of $ L  \langle m^{2} \rangle $ versus $T^{-1}$.
The slope of the solid line (linear fit to the data) is $a=1.6$.  
(b) Plots of $\langle |m| \rangle $ versus $T$.
More details in the text.}
\label{fig2}
\end{figure}

Decreasing temperature the order parameter probability 
distribution broadens, it becomes non-Gaussian, and near $T_{c}$ it
splits into two peaks that get the more separated the lower the
temperature. For $T < T_{c}$ and linear dimensions $L$ much larger
than the correlation length $\xi $ of order parameter fluctuations,
one may approximate $P_{L}(M)$ near the peaks by a double-Gaussian
distribution, i.e.
\begin{equation}
P_{L}(M) \propto \exp \left( {-(M - M_{sp})^{2}L^{d}} \over {2T \chi} \right) +
\exp \left( {-(M + M_{sp})^{2}L^{d}} \over {2T \chi} \right) \ \ ,   
\end{equation}                          
where $M_{sp}$ is the spontaneous magnetization, while
the susceptibility $\chi$ is now given by 
\begin{equation}         
\chi = \frac{L^{d}}{T} \left(\langle M^{2}\rangle - \langle |M| \rangle ^{2} \right) \ \ .
\end{equation}
From equation (4) it turns out that 
the Gaussian squared width $\sigma^{2}$
associated with high temperature distributions is very close to
the 2nd moment of the order parameter, i.e.
\begin{equation}  
\sigma^{2} \approx \langle M^{2} \rangle \ \ .
\end{equation}  
Equation (8) is a consequence of the Gaussian shape of 
the order parameter probability
distribution and, thus, it holds for the MEM as well. 
From the known one-dimensional exact solution for a
chain of $L$ spins \cite{delta,eps} one can obtain
\begin{equation}   
\chi = \frac {1}{T} \exp (2/T) \ \;
\end{equation}   
then, equations (5) and (9) lead us to 
\begin{equation}  
\langle M^{2} \rangle = \frac {1}{L} \exp (2/T) \ \          
\end{equation}  
(where it has been taken into account that
$\langle M \rangle = 0$ due to finite-size effects, irrespective
of temperature).
From equations (8) and (10) we can see that the high
temperature Gaussian probability distribution broadens exponentially
as $T$ gets lowered, until it develops delta-like peaks at $M= \pm 1$ 
as a consequence
of a boundary effect on the widely extended distribution.
It should be noticed that for $d \geq 2$ this phenomenon 
is prevented by the finite
critical temperature which splits the Gaussian, as implied by equation (6).

Turning back to the MEM, figure 1 strongly suggests that an analogous
mechanism should be responsible for the thermal dependence exhibited by
the MEM's order parameter distribution function. 
So, by analogy to equation (9), we assume the relation
\begin{equation}   
\chi = \frac {1}{T} \exp (a/T) \ \
\end{equation}   
to hold for the MEM, where we have introduced a
phenomenological parameter $a$,
and the susceptibility $\chi$ is given by equation (5). We find an excellent
agreement to the data by choosing the value $a=1.6$ as
observed in figure 2(a), where ln-linear plots
of $ L \langle m^2 \rangle$ versus $1/T$ are shown for
strip widths varying in the range $16 \leq L \leq 1024$.
Figure 2(b) shows plots of $\langle |m|  \rangle$ versus $T$ 
for the same lattices.
This figure shows that increasing $L$
the order parameter curves approach the one that corresponds to
the thermodynamic limit (i.e., $\langle |m| \rangle = \theta(T)$, 
where $\theta$ is the Heaviside function).

\begin{figure}
\centerline{{\epsfxsize=4.0in\epsfysize=3.0in \epsffile{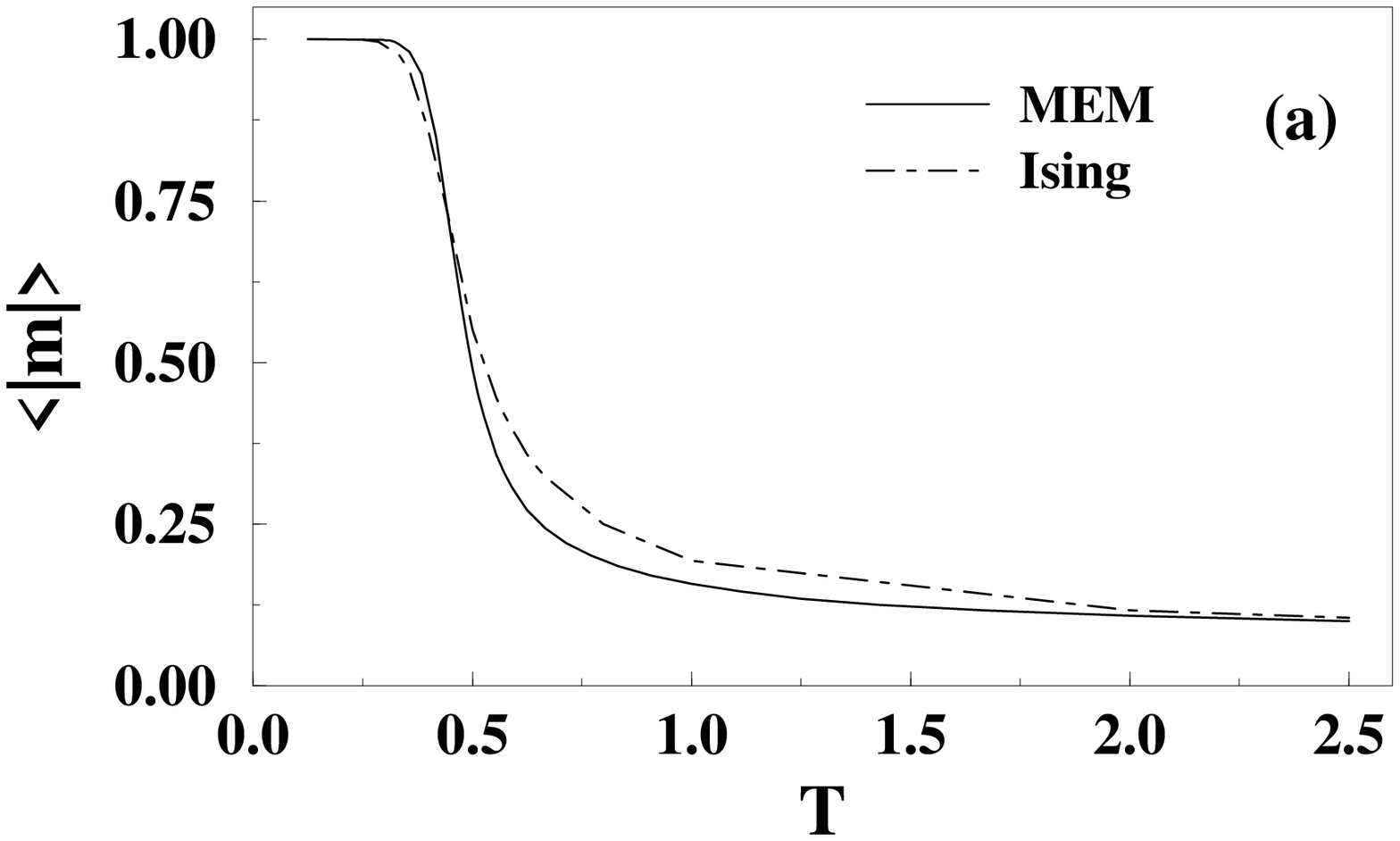}}}
\centerline{{\epsfxsize=4.0in\epsfysize=3.0in \epsffile{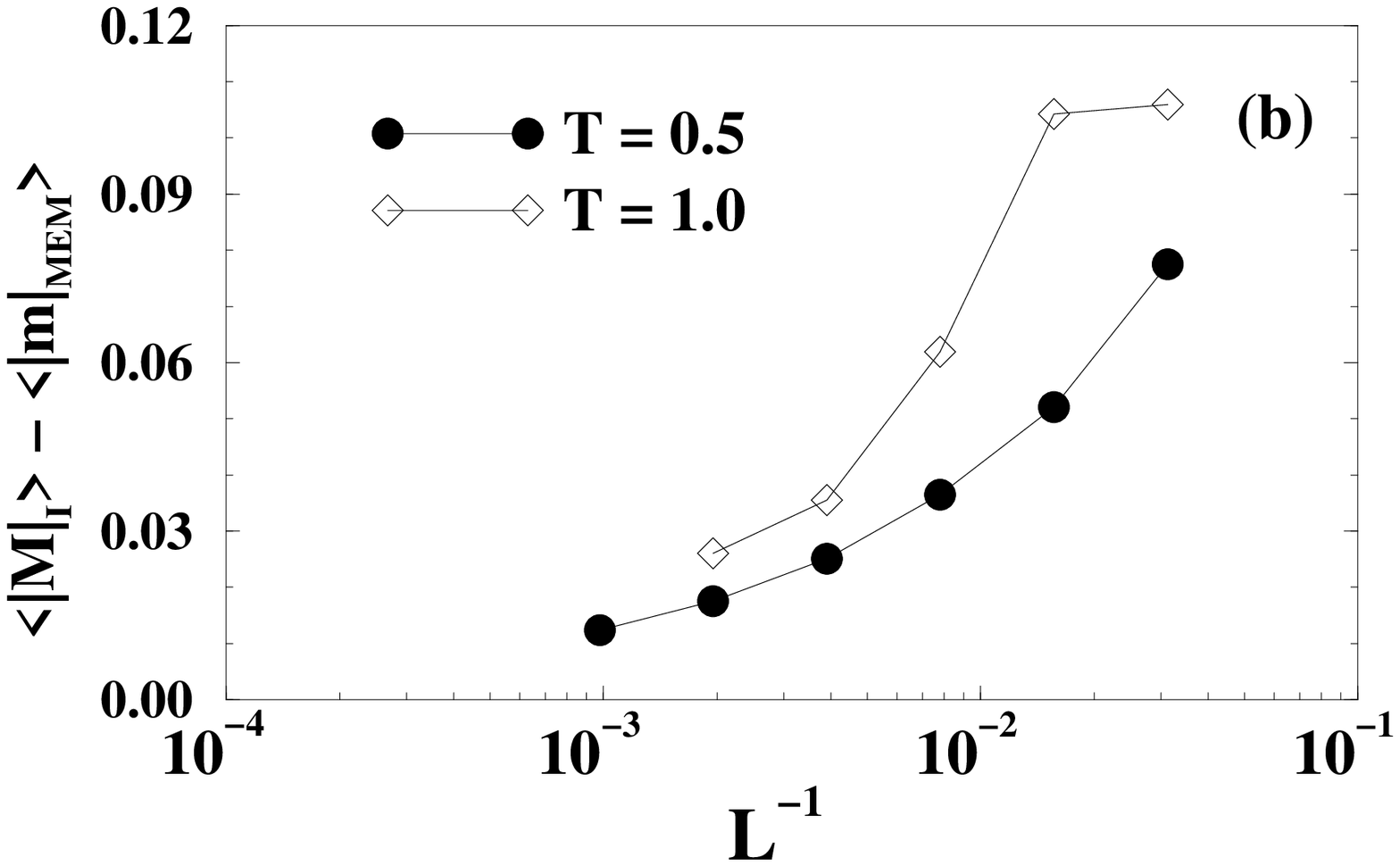}}}
\caption{Comparison of results corresponding to the 
$(1+1)-$MEM and the $d=1$ Ising model. 
(a) Plots of $\langle |m| \rangle $ versus $T$ obtained for a lattice of side $L=128$. 
(b) Linear-log plots of $\langle |M|_{I} \rangle (L,T) - \langle |m|_{MEM}
\rangle (L,T)$  versus $L^{-1}$
for $T=0.5$ and $T=1.0$. Hence, differences in the magnetization 
due to  finite-size effects appear to
vanish in the thermodynamic limit.}
\label{fig3}
\end{figure}

However, it should be pointed out that the
results obtained for the $(1+1)$-MEM and the $1$-dimensional Ising
model do not exactly coincide for finite lattices, 
as figure 3(a) shows for the case of the magnetization. Anyway,
this fact should not alarm us, since it can be seen that 
differences in the results obtained for both models are a direct 
consequence of the
finite-size nature of the lattices used in the simulations 
and consequently they tend to vanish in the
thermodynamic limit. This is actually shown by figure 3(b),
where log-linear plots of
$ \langle |M|_{Ising} \rangle (L,T) -  \langle |m|_{MEM}  \rangle (L,T)$ 
versus $L^{-1}$
for two different fixed values of temperature are presented.
Thus, we conclude that in view of the full qualitative and
quantitative agreement between both models we can safely
establish that, as in the 1d Ising model, 
the $(1+1)$-MEM is not critical (i.e. it also undergoes a
phase transition at $T_{c} = 0$).

\begin{figure}
\centerline{{\epsfxsize=4.0in\epsfysize=3.0in \epsffile{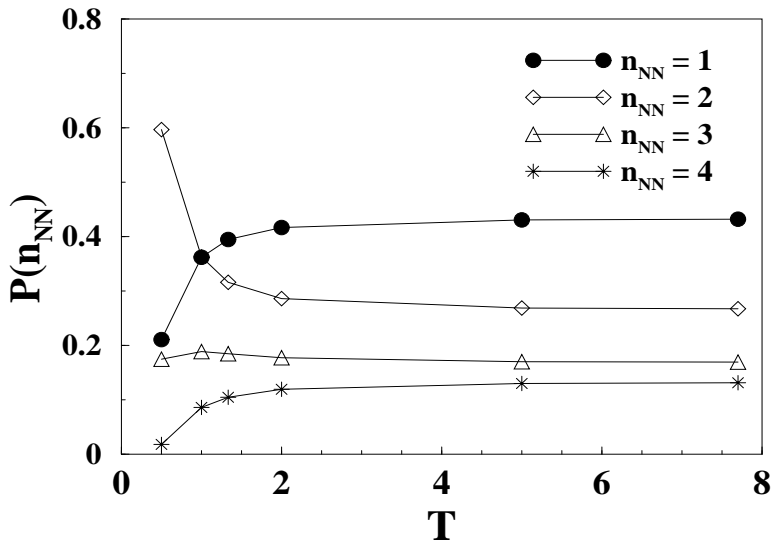}}}
\caption {Plots of $P(n_{NN})$ versus $T$ for $n_{NN}=1,2,3,4$ 
as indicated in the figure. 
The lines are guides to the eye. More details in the text.}
\label{fig4}
\end{figure}

We have also computed the number of already occupied NN 
sites every time that a
new particle was added to the spin system, and thus we have
obtained the normalized probability $P(n_{NN})$ of having $n_{NN}$ occupied
NN sites. Figure 4 shows the behavior of $P(n_{NN})$ as a function of
temperature. Using this probability we have evaluated 
$\langle n_{NN} \rangle = 2.0000(1)$ irrespective of the temperature.
This result can be understood considering 
that the growing process that leads to the assignment of a spin
$S_{ij} = \pm 1$ to each lattice site of coordinates $(i,j)$ can
be studied by means of a bond model. In fact, we can assign
a bond to each pair of neighboring sites, pointing  from the
earlier occupied site to the later occupied one. 
So, the process that leads
to a given spin configuration can be specified by the fields
$b_{U}(i,j)$ and $b_{R}(i,j)$, where sub-indexes $U$ and $R$ 
refer to the upper bond
of $(i,j)$ (i.e., the bond that connects the site of coordinates
$(i,j)$ with that of coordinates $(i,j+1)$),
and  to the right bond of the site of coordinates $(i,j)$, 
respectively \cite{note}. 
We take $b(i,j) = +1$ if the bond points outwards and 
$b(i,j) = -1$ if it is directed inwards. Therefore, 
the net bond flux at
a given lattice site $(i,j)$ is given by:
\begin{equation}  
\phi(i,j) = b_{U}(i,j) + b_{R}(i,j) - b_{U}(i,j-1) - b_{R}(i-1,j)      
\end{equation}  
and the possible 
values that $\phi$ can take are $\phi = -4, -2, 0, 2$. 
After some algebra, it follows that
$n_{NN} = \frac{1}{2} (4 - \phi)$
holds for every site on the lattice. Moreover, it can be seen that,
for an arbitrary $d$-dimensional lattice of coordination number $q$,
$n_{NN} = \frac{1}{2} (q - \phi)$. Then,
\begin{equation}  
\langle n_{NN} \rangle = \frac{1}{2} \ \ q                                  
\end{equation}
is the mean number of occupied NN, since $\langle \phi \rangle = 0$.
For the two-dimensional square lattice, $q = 4$ and equation (13) yields
$\langle {n_{NN}} \rangle = 2$,
in agreement with the result we have already obtained by means
of Monte Carlo simulations.

\begin{figure}
\centerline{{\epsfxsize=4.0in\epsfysize=3.0in \epsffile{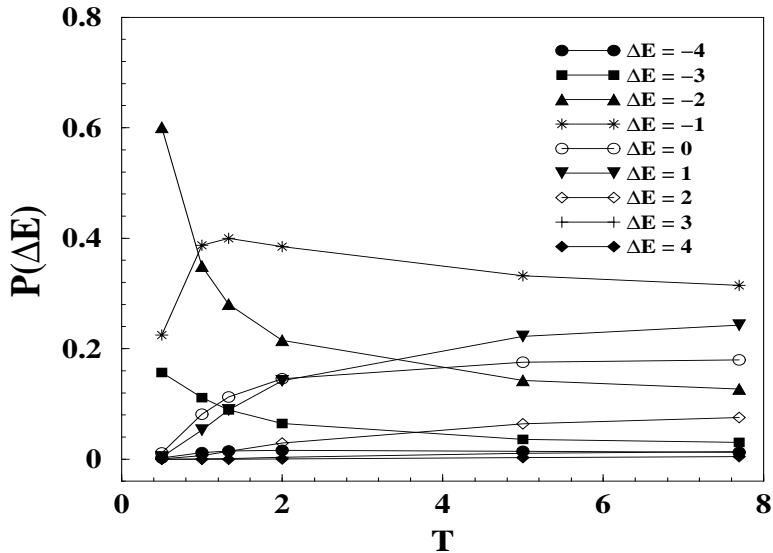}}}
\caption{Plots of $P(\Delta E)$ versus $T$ for $\Delta E = 
0, \pm 1, \pm 2, \pm 3, \pm4$ (in units of $J$) as indicated in the 
figure. The lines are guides to the eye. More details in the text.}
\label{fig5}
\end{figure}

Further insights into the MEM's growing process can be
gained by studying the mean energy change involved in the addition of
a new particle to the system. The process of adding a new spin
involves an energy change $\Delta E$ and from the definition 
of the $(1+1)-$MEM,
the possible values that $\Delta E$ can take are 
$0, \pm 1, \pm 2, \pm 3, \pm4$ (in units of $J$). Figure 5 shows 
plots of the normalized probability $P(\Delta E)$ versus $T$ 
for each of these values. 
The non-equilibrium nature of the MEM manifests itself
through much more complex probability distributions
$P(\Delta E)$ (see figure 5) 
than the corresponding to the equilibrium Ising model
where $\Delta E$ can take only five different values, namely
$0, \pm 2, \pm4$ (in units of $J$).
The results shown in  figures 4 and 5  
confirm the non-trivial nature of the link established between the MEM at
stationarity and the Ising model in equilibrium.

It should be noticed that for the case of crystal growth models (CGM)
\cite{x,y,m} the growing conditions are quite different than those of
the MEM. In fact, in CGM the crystal grows layer by layer   
in a given direction \cite{m,w}. Furthermore, the probability
distribution of the predecessor spin layer is sampled 
from the {\bf equilibrium} distribution, so will be the probability
of spins in subsequent layers. This particular growth mechanism
allow to establish dual transformations with the kinetic Ising model
\cite{d,w} and to extract some exact results. In contrast, the growing
interface of the MEM is self-affine and the system is far from equilibrium.
So, the link between the 1d Ising model and the $(1 + 1)-$dimensional
MEM is quite challenging. 

\section{Study of the MEM in $(2+1)-$dimensions: results and discussion}

\subsection{The order parameter and its probability distribution
 function}

\begin{figure}
\centerline{{\epsfxsize=4.0in\epsfysize=3.1in \epsffile{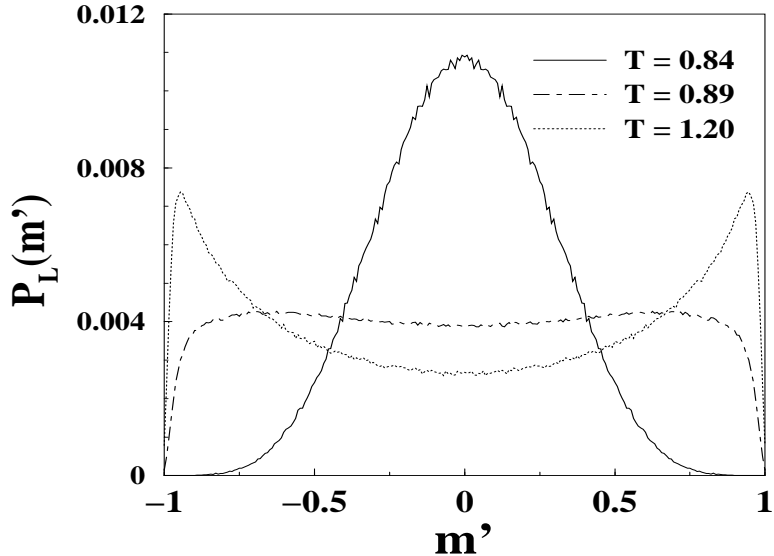}}}
\caption{Plots of the probability distribution 
$P_{L}(m^{'})$ versus $m^{'}$ for the fixed lattice
size $L=16$ and different temperatures, as indicated in
the figure.
The occurrence of two maxima located at 
$m^{'} = \pm M_{sp}$ (for a given value of $M_{sp}$
such that $0 < M_{sp} < 1$)
is the hallmark of a thermal continuous phase transition
that takes place at a finite critical temperature.}
\label{fig6}
\end{figure}

In order to compare the $(2 + 1)-$dimensional MEM and the 2d Ising model,
we have first studied the order parameter probability
distribution $P_{L}(m^{'})$, where $m^{'}$ takes now $L^{2}+1$ 
possible values (see figure 6).
For high temperatures, the probability distribution
corresponds to a Gaussian centered at $m^{'}=0$. At lower
temperatures we observe the onset of two
maxima located at $m^{'} = \pm M_{sp}$ $(0 < M_{sp} < 1)$,
which become sharper and approach 
$m^{'} = \pm 1$ as $T$ is gradually decreased.

\begin{figure}
\centerline{{\epsfxsize=4.0in\epsfysize=3.0in \epsffile{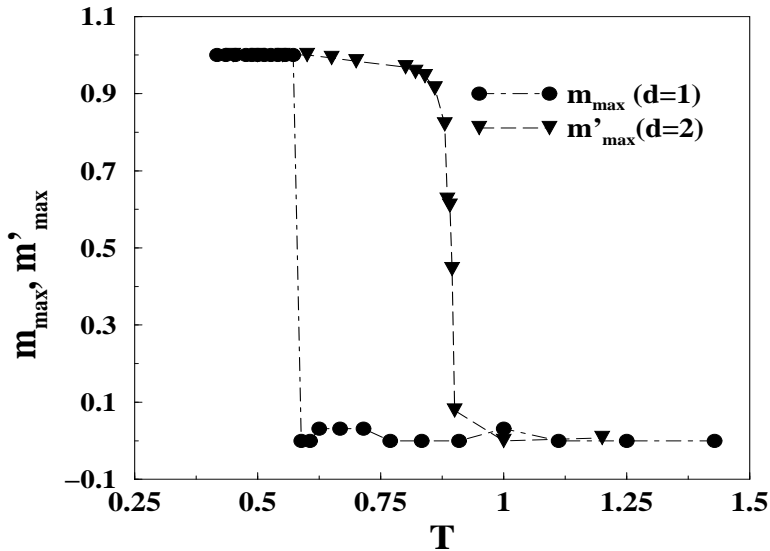}}}
\caption{Plots showing
the location of the maximum of the probability distribution
as a function of temperature for both 
$(d+1)-$dimensional MEM models ($d=1,2$). The lines are guides to the eye.
The smooth transition
for d=2 constitutes another evidence of the
finite critical point associated with the $(2+1)-$MEM.}
\label{fig7}
\end{figure}

Figure 7 shows
the location of the maximum of the probability distribution
as a function of temperature for both 
$(d+1)$-dimensional MEM models, with  $d=1,2$.
While for the $d=2$ case we observe a smooth
transition from the $m^{'}_{max}=0$ value characteristic of high temperatures
to nonzero $m^{'}_{max}$ values that correspond to lower temperatures,
the curve obtained for $d=1$ shows, in contrast, a Heaviside-like jump.
As already discussed, the behavior exhibited by the
$(2+1)$-dimensional MEM 
(e.g., as displayed by figures 6 and 7) is the signature
of a thermal continuous phase transition that takes place at
a finite critical temperature.

The broken symmetry at a finite critical temperature $T_{c}$ implied by
the thermal continuous phase transition can be explained in terms of the
broken ergodicity that occurs in the system when we tend to
the thermodynamic limit ($L \rightarrow \infty$)
making use of the temperature dependence exhibited by the
order parameter distribution function.
In fact, if we set the characteristic length of MEM's
domains $l_{D}$ equal to an ergodic length $l_{erg}$,
we can carry out a complete analogy with the Ising model
by associating $l_{erg}$ to $t_{erg}$ (the Ising model ergodic time)
and the cluster's total length $l_{C}$ (already defined in Section 3)
to the Ising model observation time $t_{obs}$.
In this way, we encounter that
excursions of $m^{'}$ from $m^{'} = +M_{sp}$ to $m^{'} = -M_{sp}$ and
{\it vice versa} occur at
length scales of the order of  $l_{erg}$.
When the cluster's
total length becomes larger and larger ($l_{C} \gg l_{erg}$) the whole
cluster's magnetization is averaged to zero.
Furthermore, $l_{erg}$ diverges
as the strip's width becomes larger and larger, and again broken
symmetry arises as the consequence of broken ergodicity.

\subsection{Order-disorder phase transition in the $(2+1)$-dimensional MEM:
Finite-size effects and scaling analysis}

As already anticipated and as it follows from figures 6 and 7,
the $(2+1)$-dimensional MEM exhibits a 
thermally driven order-disorder transition at a finite temperature.
In the thermodynamic limit ($L \rightarrow \infty$) 
we expect to determine a
critical temperature $T_{c}$ such that $\langle |m^{'}| \rangle =0$ for $T > T_{c}$ 
while $\langle |m^{'}| \rangle$ remain non-vanishing at temperatures below $T_{c}$.

From the finite-size
scaling theory, developed for the treatment of
finite-size effects at criticality and under 
equilibrium conditions \cite{barba,priv}, 
it is well known that if a thermally driven
phase transition occurs at a temperature $T_{c} > 0$ in the
thermodynamic limit, then in a confined geometry this
transition becomes smeared out over the temperature region
$\Delta T(L)$ around a shifted effective transition temperature
$T_{c}(L)$, and the following relationships hold:
\begin{equation} 
\Delta T(L) \propto L^{-\theta}   \ \ ,
\end{equation} 
and 
\begin{equation}              
|T_{c}(L)-T_{c}| \propto  L^{-\lambda} \ \  ,
\end{equation}         
where the rounding and shift exponents are given by
$\theta = \lambda = \nu^{-1}$, respectively (recalling
that $\nu$ is the exponent that characterizes the divergence
of the correlation length at criticality).

Furthermore, from well established finite-size scaling relations,
the following Ans\"atze hold just at criticality $(T = T_{c})$ :
\begin{equation} 
\langle |m^{'}(L,T = T_{c})| \rangle \propto L^{-\beta/ \nu} 
\end{equation}
and
\begin{equation}                  
\chi_{max}(L)  \propto L^{\gamma/ \nu} \ \ ,   
\end{equation} 
where $\beta$ and $\gamma$ are the order
parameter and the susceptibility critical exponents,
respectively. Note that $\chi_{max}(L)$, as given by equation (17),
refers to the maximum of $\chi(L,T)$ as a function of $T$ for
fixed lattice size $L$.

\begin{figure}
\centerline{{\epsfxsize=4.0in\epsfysize=3.0in \epsffile{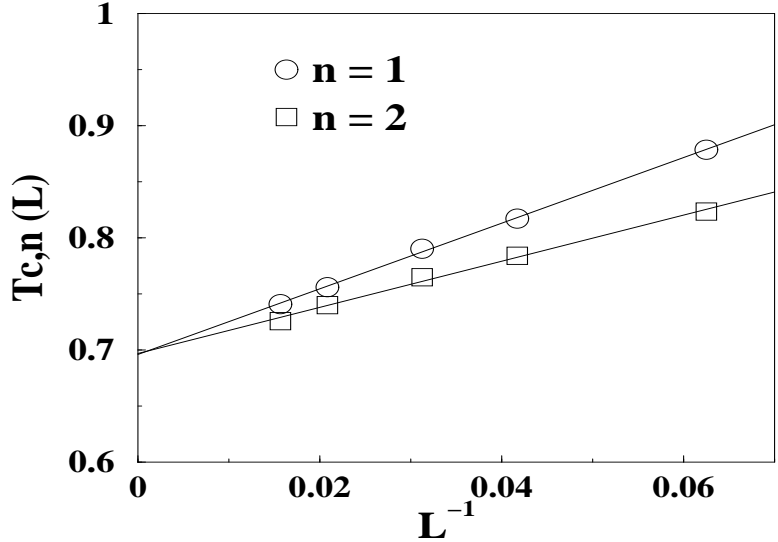}}}
\caption {Plots of $T_{cn}(L)$ versus $L^{-1}$ (for $n=1,2$). The
solid lines show the linear extrapolations that meet at the
critical point given by $T_{c}=0.69 \pm 0.01$.}
\label{fig8}
\end{figure}
 
In view of the encountered analogies between the MEM and the 
Ising model, it is natural to test the validity of equations 
(14-17) for the case of the MEM in $(2 + 1)$-dimensions. 
It should be noted that as in the case of equilibrium systems, 
in the present  case various ``effective" L-dependent critical 
temperatures can also be defined. In particular, we will define $T_{c1}(L)$ 
as the value that corresponds to $\langle |m^{'}| \rangle = 0.5$ for fixed $L$,
and $T_{c2}(L)$ as the one corresponding to the maximum of the
susceptibility for a given $L$, assuming that the susceptibility
is related to order parameter fluctuations in the same manner
as for equilibrium systems (as given by equations (5) and (7)).
Then, we should be able to obtain $T_{c}$ from
plots of $T_{cn}(L)$ versus $L^{-1}$  (for $n=1,2$),
as it is shown in figure 8.
Following this procedure we find that both $T_{c1}(L)$ and $T_{c2}(L)$
extrapolate (approximately) to the same value, allowing us 
to evaluate the critical temperature
$T_{c} = 0.69 \pm 0.01 $ in the thermodynamic limit. 

After determining $T_{c}$,
the correlation length exponent $\nu$
can be evaluated by means of equation (15), making the replacement
$\lambda = 1/ \nu$. In fact, taking $T_c$ 
at the mean, maximum and minimum
values allowed by the error bars, we obtain six log-log plots
of $|T_{cn}(L)-T_{c}|$ versus $L$ for $n=1,2$.
The slope of each of these plots, not shown here for the sake of space,
yields a value for $\nu$. The obtained values are:
$$
        \nu = 1.08 (T_{c}=0.68),\ \
        \nu = 1.00 (T_{c}=0.69),\ \ 
$$
\begin{equation} 
        \nu = 0.88 (T_{c}=0.70)\ \ \rm{for} \ \ n=1,
\end{equation}
and
$$
        \nu = 1.20 (T_{c}=0.68),\ \
        \nu = 1.08 (T_{c}=0.69),\ \ 
$$
\begin{equation}
        \nu = 0.95 (T_{c}=0.70)\ \  \rm{for} \ \ n=2.
\end{equation} 

Thus our estimate is given by $\nu = 1.04 \pm 0.16 $,
where the error bars reflect the
error derived from the evaluation of $T_{c}$, as well as the
statistical error.

\begin{figure}
\centerline{{\epsfxsize=4.0in\epsfysize=3.0in \epsffile{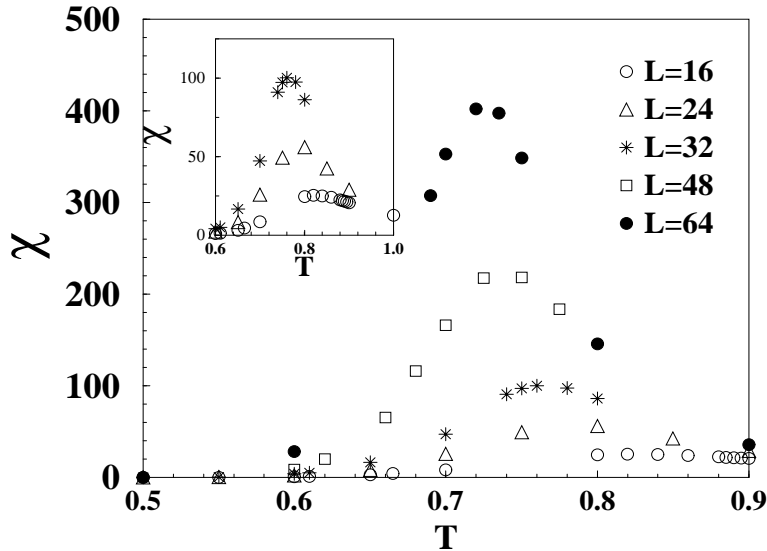}}}
\caption{Behavior of the susceptibility as a function of temperature.
Each curve shows a peak
which becomes sharper and shifts towards lower
temperatures as $L$ is increased. 
The inset shows in greater detail the peaks
corresponding to the smaller lattices ($L=16,24,32$).} 
\label{fig9}
\end{figure}

Figure 9 shows plots of the susceptibility versus $T$ as 
obtained using lattices of different side. It is found that the susceptibility
exhibits a peak which becomes sharper and shifts towards lower
temperatures when $L$ is increased. 
This behavior is, in fact,
already anticipated by equation (17), and it allows us to evaluate
$\gamma/\nu$ from the slope of a log-log plot of
$\chi_{max}$ versus $L$,
as figure 10 shows. The linear fit yields $\gamma/\nu = 2.02 \pm 0.04$.
Using this value and the value formerly obtained for $\nu$ we thus
determine $\gamma= 2.10 \pm 0.36$.

\begin{figure}
\centerline{{\epsfxsize=4.0in\epsfysize=3.0in \epsffile{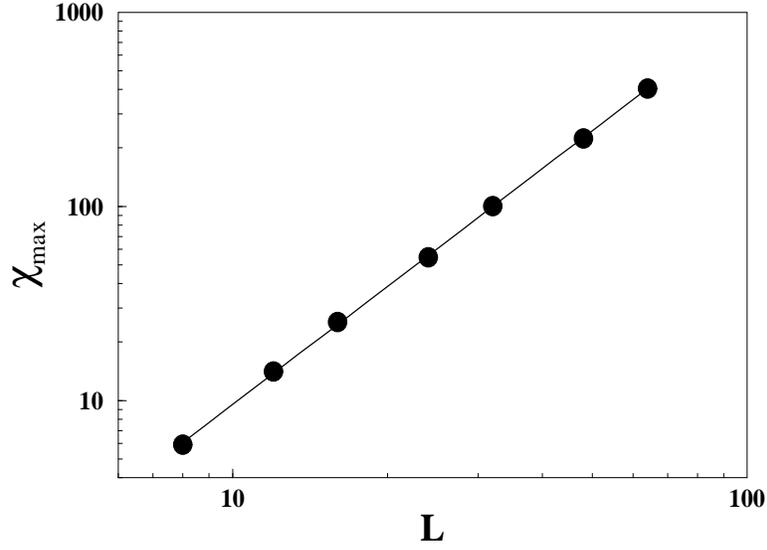}}}
\caption {Log-log plot of $\chi_{max}$ versus $L$.
The linear fit (solid line) yields $\gamma/\nu = 2.02 \pm 0.04$.}
\label{fig10}
\end{figure}

\begin{figure}
\centerline{{\epsfxsize=4.0in\epsfysize=3.0in \epsffile{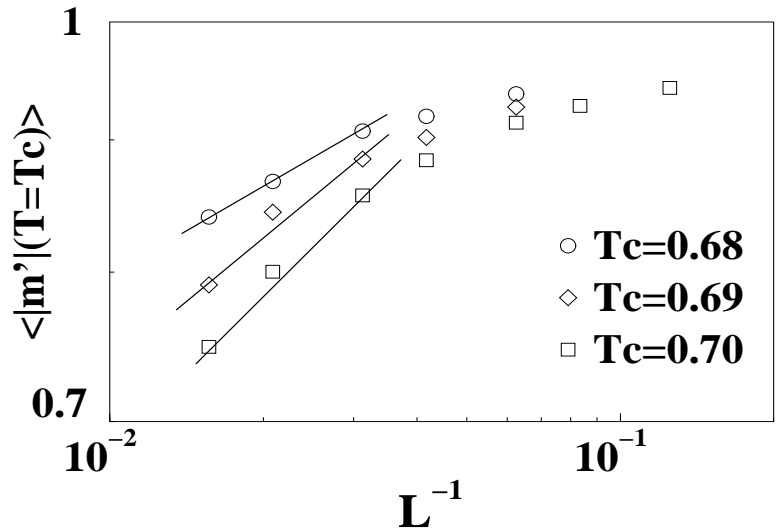}}}
\caption{Log-log plots of 
$\langle m^{'}(T=T_{c}) \rangle$ versus $L^{-1}$ for the
mean, maximum and minimum allowed values of $T_{c}$.
The linear fits (solid lines) yield an estimate 
$\beta/\nu = 0.16 \pm 0.05$.}
\label{fig11}
\end{figure}

Figure 11 shows log-log plots of $\langle |m^{'}| \rangle(T=T_{c})$ versus $L$
for the mean, maximum and minimum allowed values of $T_{c}$.
Considering only the larger lattices, the linear fits to the data  
according to equation (16) yield the following estimates: $\beta/\nu = 0.11$, 
$\beta/\nu = 0.16$ and  $\beta/\nu = 0.19$. We then assume the value 
$\beta/\nu = 0.15  \pm 0.04$, where the error bars reflect the
error derived from the evaluation of $T_{c}$, as well as the
statistical error. From this value and the value formerly obtained 
for $\nu$ we thus determine $\beta= 0.16 \pm 0.05$. 
 
The critical exponents of the MEM in $(2 + 1)-$dimensions
as obtained using a finite-size scaling analysis are, so far:
$\nu = 1.04 \pm 0.16 $, $\gamma = 2.10 \pm 0.36$ and
$\beta = 0.16 \pm 0.05$.
If we recall the exactly known critical exponents of 
the $d = 2$ Ising model, i.e. $\nu = 1$, $\gamma = 7/4$ and
$\beta = 1/8$, we find that the $(2+1)-$MEM has the same 
critical exponents within error bars.
These results further support our conjecture on the connection between
the MEM in $(2 + 1)$-dimensions and the Ising model in $2$ dimensions.

As in the case of the MEM in $d = 1$, we have also computed 
the number of already occupied NN sites every time that a
new particle was added to the spin system. 
We found that the value $\langle n_{NN} \rangle = 3.0000(1)$ holds for all
temperatures, which is indeed the result given by equation (13),
since $q = 6$ for the three-dimensional square lattice.

At this stage, we may recall that for the 
$(1+1)-$MEM $\langle n_{NN} \rangle$ equals
the coordination number of the $d=1$ Ising model,
and that we found that both models
have the same critical temperature and exhibit the same critical behavior.
Reasoning by analogy, we may expect a coincidence between the critical
temperature for the $(2+1)-$MEM and the corresponding one for a
$d=2$ Ising model defined on a lattice of coordination number
$q=3$. However, this comparison cannot be carried out, since the
critical temperature of an Ising model depends on both
the coordination number $q$ and the topological structure of the lattice,
but for $d \geq 2$ and a given value of $q$ the
topological structure is not unique.
For instance, for $d=2$ and $q=3$, we can pass from the honeycomb lattice (HL)
to the expanded Kagom\'{e} lattice (EKL) through the application 
of a star-triangle
transformation and obtain the exact values of their critical points,
which turn out to be \cite{fisher} 
$ T_{c} = 1.5187$ (HL) and $ T_{c} = 1.4530$ (EKL).

\section{Conclusions}

In the present work we have studied the growth of magnetic
Eden clusters with ferromagnetic interactions between
nearest neighbor spins in a $(d+1)-$dimensional rectangular
geometry (for $d=1,2$), using Monte Carlo simulations and applying a
finite-size scaling theory.
The results obtained allow us to conjecture
a nontrivial correspondence between the MEM for the irreversible
growth of magnetic materials and the classical Ising model
under equilibrium conditions. In fact, we have 
found that the $(d+1)-$dimensional
MEM and the $d-$dimensional Ising model behave identically 
(unless finite-size differences that vanish in the
thermodynamic limit) at criticality,
i.e. that both models belong to the same universality class.
We also conjecture that
this correspondence would remain at higher dimensions $(d>2)$. 
The results obtained strongly suggest a
link between the temporal evolution of equilibrium systems and
the stationary growth of nonequilibrium systems.
We thus believe that this work will stimulate further developments
in the field of nonequilibrium kinetic growth models. A more 
precise numerical test of the posed conjecture will certainly 
require a considerable computational effort but it will be 
of great interest.  Furthermore, analytical
developments aimed to establish a theoretical framework
for the understanding of far from equilibrium growth phenomena
will become stimulated by the reported findings.

\section*{Acknowledgments} This work was supported  financially  by
CONICET, UNLP, CIC (Bs. As.), ANPCyT and Fundaci\'on Antorchas 
(Argentina) and the Volkswagen Foundation (Germany). The authors
thank M. Mu\~noz for helpful discussions and his 
proposal of the bond representation for the MEM.


\begin{thebibliography}{99}
\bibitem{note} Since we are dealing with periodic boundary conditions in
the $j$-direction, we consider $j-1 = L$ if $j=1$ and analogously 
$j+1 = 1$ if $j=L$ throughout.
\end{thebibliography}
\end{document}